\documentclass[sigconf]{acmart}

\usepackage{xcolor,xspace,multirow,hyperref}
\AtBeginDocument{%
\providecommand\BibTeX{{%
\normalfont B\kern-0.5em{\scshape i\kern-0.25em b}\kern-0.8em\TeX}}}
\usepackage{wrapfig}

\newcommand{\msec}[1]{\S\,\ref{#1}}

\usepackage{multirow}
\newcommand{\minitab}[2][l]{\begin{tabular}{#1}#2\end{tabular}}

\newcommand{\add}[1]{{#1}}
\newcommand{\rev}[2]{{#2}}

\copyrightyear{2024}
\acmYear{2024}
\setcopyright{acmlicensed}\acmConference[CHI '24]{Proceedings of the CHI Conference on Human Factors in Computing Systems}{May 11--16, 2024}{Honolulu, HI, USA}
\acmBooktitle{Proceedings of the CHI Conference on Human Factors in Computing Systems (CHI '24), May 11--16, 2024, Honolulu, HI, USA}
\acmDOI{10.1145/3613904.3642160}
\acmISBN{979-8-4007-0330-0/24/05}

\begin{document}

\title{Generative AI in the Wild: \rev{Marvelous, Mysterious, and Maddening}{Prospects, Challenges, and Strategies}}



\author {Yuan Sun}
\email{yuan.sun@ufl.edu}
\orcid {0000-0002-0752-1402}
\affiliation{%
  \institution{University of Florida}
  {Gainesville, FL}
  \country{USA}
  }

\author {Eunchae Jang}
\email{ezj5160@psu.edu}
\orcid{0000-0001-7547-9050}
\affiliation{%
  \institution{The Pennsylvania State University}
  \city{University Park, PA}
  \country{USA}
  }

\author {Fenglong Ma}
\email{fenglong@psu.edu}
\orcid{0000-0002-4999-0303}
\affiliation{%
  \institution{The Pennsylvania State University}
  \city{University Park, PA}
  \country{USA}
  }

   \author {Ting Wang}
   \email{twang@cs.stonybrook.edu}
   \orcid{0000-0003-4927-5833}
   \affiliation{%
  \institution{Stony Brook University}
  \city{Stony Brook, NY}
  \country{USA}
  }

\begin{abstract}

Propelled by their remarkable capabilities to generate novel and engaging content, Generative Artificial Intelligence (GenAI) technologies are disrupting traditional workflows in many industries. While prior research has examined GenAI from a techno-centric perspective, there is still a lack of understanding about how users perceive and utilize GenAI in real-world scenarios. To bridge this gap, we conducted semi-structured interviews with ($N$ = 18) GenAI users in creative industries, investigating the human-GenAI co-creation process within a holistic \rev{preparation--process--product}{LUA (\underline{L}earning, \underline{U}sing and \underline{A}ssessing)} framework. Our study uncovered an intriguingly complex landscape: \rev{Marvelous--}{Prospects --} GenAI greatly fosters the co-creation between human expertise and GenAI capabilities, profoundly transforming creative workflows; \add{Challenges -- Meanwhile, users face substantial uncertainties and complexities arising from resource availability, tool usability, and regulatory compliance}; \rev{Maddening} {Strategies -- In response, users actively devise various strategies to overcome many of such challenges}. Our study reveals key implications for the design of future GenAI tools. 
\end{abstract}

\ccsdesc[500]{Human-centered computing~User studies}

\keywords{Generative AI, Human-AI Collaboration, Transparency, User Agency}

\maketitle

\section{Introduction}

Generative Artificial Intelligence (GenAI), the latest advances in AI technologies, is designed to generate novel content, insights, and solutions by identifying, replicating, and recomposing intricate patterns within existing data. 
In the past few years, GenAI has been transforming a spectrum of industries in a disruptive and profound way, enabling use cases previously considered strictly experimental, ranging from generating realistic visual arts~\cite{genai-art}, conducting human-like conversations~\cite{zamfirescu2023johnny}, and writing programs~\cite{genai-coding}, to even inventing new medical treatments~\cite{genai-bio}. 

However, despite its immense potential, the current discourse around GenAI is a blend of optimism and discouragement. On one hand, GenAI is envisioned to unlock unprecedented avenues for efficiency, creativity, and productivity~\cite{genai-at-work}. On the other hand, incorporating GenAI tools into concrete domains often faces major challenges~\cite{chatgpt-challenge}, resulting in limited adoption of GenAI by practitioners~\cite{genai-adoption}. \add{Unlike conventional AI that focuses on analysis, decision-making, or automation, GenAI specializes in
creating new content that does not exist before. Thus, while existing literature studies how users perceive and use conventional AI technologies~\cite{kocielnik2019will,gursoy2019consumers}, the unique characteristics of GenAI (e.g., its creativity) have yet to be fully understood from a human-centric perspective.} In addition, recent HCI studies have started investigating how GenAI tools are being used for specific tasks, such as creative writing~\cite{gero2023social, yuan2022wordcraft}, music composition~\cite{genai-music}, and chatbot building~\cite{zamfirescu2023johnny} and exploring effective functions to improve user experience~\cite{dang2023choice, wu2022aichains, wang2023reprompt}. However, in real-world scenarios, the concrete tasks are defined and executed by users with varying roles, interests, and backgrounds~\cite{passi2018trust}. Users often face a variety of challenges beyond GenAI's functionality, define their goals according to complex situations, and use varied strategies to accomplish such goals~\cite{suchman1987situated}. 
Thus, \rev{using the isolated task context as a scaffold to achieve GenAI effectiveness may limit our understanding of how users perceive and use GenAI in practical environments, while relying primarily on a techno-centric perspective may fail to}{it is crucial to understand how GenAI is integrated into users' workflows by} accounting for the `social nuances, affective relationships, or ethical, value-driven concerns' of GenAI users~\cite{aragon2016developing}. \add{This necessitates a transition from viewing human-GenAI interaction as a static, single instance of tool usage to understanding it as a dynamic, evolving process.}.

\begin{figure*}[!t]
\centering
\includegraphics[width=0.8\textwidth]{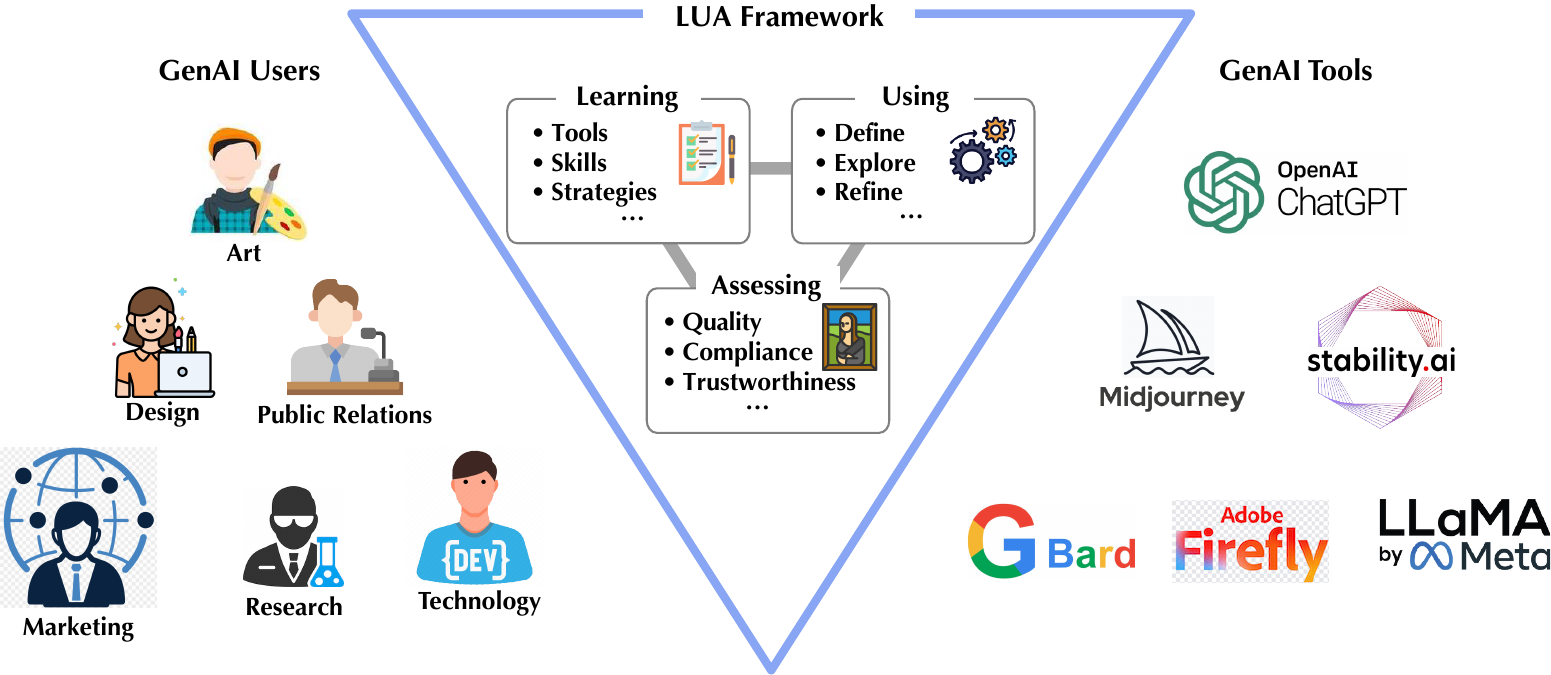}
\caption{\add{Co-creation between human expertise and GenAI capabilities in complex, real-world settings.}}
\label{fig:my_label}
\end{figure*}

The present study aims to fill this critical research gap by understanding how users perceive and utilize GenAI \rev{in real-world settings}{`in the wild.'} \add{
While GenAI is poised to disrupt many sectors of our society, we focus our study on creative industries, professions that generate original ideas, concepts, or content for various forms of media, entertainment, design, or communication~\cite{foord2009creativeindustries}. Our rationale is as follows. {\em i}) Creative industries have emerged as one front-runner in adopting GenAI technologies~\cite{anantrasirichai2022artificial}, which enable creative professionals to expedite and automate content generation~\cite{hughes2021generative,genai-art,muller2023genaichi23}. {\em ii}) Creative industries encompass fields such as graphic design, advertising, fashion, writing, visual arts, and many others, providing heterogeneous, diverse settings of professionals (e.g., writers, artists, and designers), content modalities (e.g., text, image, and videos), and GenAI tools (e.g., ChatGPT, Midjourney, and Stability.AI) to study real-world users' perceptions and uses of GenAI. {\em iii}) Given the critical importance of novelty and originality in content for creative industries~\cite{sakirin2023user}, we have the opportunity to thoroughly examine how users perceive GenAI's creativity. {\em iv}) As most creative professionals are neither technical experts nor policy-makers, we are able to observe how technical and non-technical challenges (e.g., ethical concerns~\cite{whoamI2023ai}) affect ordinary users' workflows.}

Thus, we conducted semi-structured interviews with ($N$ = 18) real-world users in the creative industries, varying in job roles, daily tasks, and expertise levels. By examining the human-GenAI co-creation process through the lens of a cyclic, iterative \rev{preparation–process–product}{LUA (\underline{L}earning, \underline{U}sing, and \underline{A}ssessing)} framework as illustrated in Fig.~\ref{fig:my_label}, our study uncovered an intriguingly complex landscape:
\begin{itemize}
\item \rev{Marvelous}{`Prospects'} -- GenAI greatly fosters the co-creation between human expertise and AI capabilities, reshaping the workflows of creative industries profoundly; 
\item \rev{Maddening}{`Challenges'} -- \rev{Throughout the co-creation process,}{Meanwhile,} users encounter a multitude of \add{uncertainties and complexities} related to resource availability, tool usability, and regulatory compliance.
\add{\item `Strategies' -- In response, users actively exercise user agency to overcome many of such challenges through continual learning, exploration, and experimentation.}

\end{itemize}

Our contributions to HCI literature are multi-fold. First, to our best knowledge, this work represents the first study to investigate how users perceive and utilize GenAI in real-world contexts, in which the use of GenAI is conceptualized as a \add{dynamic, evolving} process, \add{rather than merely a static, single instance of tool usage}. Second, we examine the co-creation between human expertise and AI capabilities within a cyclic, iterative \rev{preparation--process-product}{learning, using, and assessing} framework, leading to rich findings about the multifaceted nature of human-GenAI interaction \add{in the creative industries}. Third, our findings provide key implications for the design of future GenAI tools from a user-centric perspective.

\section{Related Work}

\subsection{Human-(Gen)AI Interaction}

\add{Prior research has explored various factors that may influence how users perceive, interact, and accept new AI technologies. Ostrom et al.~\cite{ostrom2019customer} examined the impact of AI-specific factors (e.g., privacy concerns, trust, and perceptions of `creepiness') besides established constructs on users' acceptance of AI in service encounters. Chiang et al.~\cite{chiang2020exploring} demonstrated that users' mental models significantly affect their assessments, expectations, and intentions regarding smart-home technologies. In a similar vein, Kocielnik et al.~\cite{kocielnik2019will} conducted empirical studies in the context of AI-enabled scheduling assistants to show that users' perceptions of accuracy and acceptance hinge on their expectations.} 
Gursoy et al.~\cite{gursoy2019consumers} developed Artificial Intelligence Device Use Acceptance (AIDUA), a three-stage acceptance model to explain users' willingness to accept AI device use in service encounters. Specifically, the study found six predictors that significantly influence customers’ AI adoption behavior, including social influence, hedonic motivation, anthropomorphism, performance expectancy, effort expectancy, and emotion toward the use of AI devices in service.

Compared with conventional AI that focuses on analysis, decision-making, or automation, GenAI specializes in creating new content that does not exist before. Thus, while similar factors (e.g., efficiency, privacy, and automation) also influence its acceptance, GenAI demonstrates many unique characteristics affecting users' perceptions. For example, Ma and Huo~\cite{ma2023users} expanded the AIDUA model and identified perceived novelty and humanness as two main factors that drive users' acceptance of ChatGPT. Similarly, in the task of creative writing, Gero et al.~\cite{gero2023social} found that users' values regarding authenticity and creativity influence their willingness to seek help from GenAI. This work extends this line of research to further investigate how these unique characteristics of GenAI, its creativity in particular, are perceived within the human-GenAI co-creation process in real-world contexts.

\subsection{\rev{Uses of GenAI}{GenAI in HCI Research}}

\add{A plethora of HCI literature has investigated how users adapt new AI technologies to their tasks and contexts. For example, Crisan and Fiore-Gartland~\cite{automl-enterprise} studied how enterprise users incorporate AutoML into their data science workflows, while You and Gui~\cite{chatbot-ai} studied how to adapt AI-enabled chatbots to support self-diagnosis. From a theoretical perspective, Russell~\cite{russell1995stages} proposed a six-stage model to describe how users adapt to new technologies, which involves awareness, learning the process, understanding and application of the process, familiarity and confidence, adaptation to other contexts, and creative application to new contexts.}

\add{In contrast to conventional AI, which requires inputting data and receiving a decision or analysis in return, GenAI typically involves iteratively providing initial inputs (e.g., prompts, styles, or specifications) and receiving novel, AI-generated outputs.}
\rev{For example, researchers have delved into understanding how users adapt GenAI tools for}{Recent work has delved into understanding how this paradigm shift impacts users' utilization of GenAI in concrete tasks such as} creative writing, music composition, and product design. Yuan et al.~\cite{yuan2022wordcraft} discovered that writing a story with GenAI enriches writing experiences in unique ways, including facilitating an interactive conversation about story topics, stimulating idea generation, and offering suggestions to aid users in surmounting creative hurdles. Copet et al.~\cite{genai-music} examined how to adapt GenAI for music composition by conditioning on textual description or melodic structure, providing users with more control over the output and allowing for a more personalized and interactive music composition experience. A recent study~\cite{zamfirescu2023johnny} explored how non-experts designed a chatbot using a large-language-model (LLM)-based tool~\cite{gpt-3}, collected user feedback regarding this specific tool, and summarized the designers' expectations for future LLM-based tools.

While most existing studies cast users' utilization of GenAI in specific tasks as a {\em static, single instance} of tool usage, it is crucial to recognize that human-GenAI co-creation in real-world contexts is often a {\em dynamic, iterative process}, in which users' understanding and use of GenAI evolve over time after learning and interacting with the tools at hand. 
The process involves skill acquisition, idea exploration, content creation, quality assessment, and solicitation of approval from various stakeholders. Thus, this work conceptualizes the use of GenAI as a dynamic, iterative process and explores how users integrate GenAI into their creative workflows from this process-oriented perspective.

\subsection{\add{GenAI ``in the Wild''}}

A growing body of HCI research has shifted its attention toward investigating how users interact with, adapt to, and integrate technological elements into real-world settings, which is often referred to as the `in-the-wild' research~\cite{chamberlain2012wild1,rogers2017wild2,kuutti2014turnwild3,rogers2011interactionwild4,rooksby2013wildsituatedactions}. \add{This in-the-wild approach emphasizes understanding how users behave, adapt, and integrate technologies into their everyday lives since ``people's understanding of the world, themselves, and interaction is strongly informed by their varying physical, historical, social, and cultural situations.''~\cite{harrison2007three} For instance, Yang et al.~\cite{yang2020re} interviewed product designers/managers to study how they incorporate or fail to incorporate AI in their products; Sun et al.~\cite{sun2023automl} examined challenges and users' workarounds of using AutoML tools in real-world settings.} 

GenAI is evolving rapidly, bringing forth a vibrant but turbulent landscape that challenges users to navigate and comprehend its potential, limitations, and implications. {In real-world settings, users tend to have varying roles, interests, and backgrounds~\cite{passi2018trust} and face a multitude of challenges beyond GenAI's functional limitations.} For example, one challenge is to determine the ownership of work co-created by humans and GenAI~\cite{fui2023generative,rezwana2023user}. Also, users may feed a considerable amount of sensitive information to GenAI tools, raising severe data privacy and security concerns~\cite{siau2020artificial}. In response, some countries~\cite{Ijaz} and organizations~\cite{Samy_2023} have prohibited the use of GenAI tools such as ChatGPT. 
Further, due to its unsupervised nature, GenAI tends to fabricate seemingly accurate but nonsensical information (i.e., hallucination)~\cite{alkaissi2023artificial, fui2023generative,genai-bias,sundar2023calling}. \add{Thus, improving GenAI's functionality (e.g., prompting)~\cite{wang2023reprompt,oppenlaender2022creativity, wu2022aichains} alone is insufficient to address users' non-functional concerns in real-world contexts.} \add{Moreover, rather than passively applying GenAI tools, users often actively exercise user agency to devise various strategies to overcome such challenges for more effectively integrating GenAI into specific, context-dependent scenarios~\cite{suchman1987situated}. Thus, understanding users' initiative is crucial for comprehending the dynamics of human-GenAI interaction.}

\add{All the above complexities create a `wild world' of GenAI in this early stage of its adoption but also present a window of opportunity for exploration. Using isolated task contexts as a scaffold may restrict our understanding of users' perceptions and utilization of GenAI in practical environments. Further, relying on a techno-centric perspective may overlook the `social nuances, affective relationships, or ethical, value-driven concerns' of GenAI users~\cite{aragon2016developing}. Therefore, we perceive the in-the-wild approach as a suitable method to guide our study.}

\section{Method}
To understand real-world users' perceptions of GenAI and how they leverage GenAI tools within their workflows, we conducted semi-structured interviews with ($N = 18$) GenAI users in the creative industries. 

\subsection{Recruitment and Interviews}
 We focused on practitioners who have hands-on experience with GenAI tools in their daily workflows. We recruited participants by spreading recruitment messages through word of mouth ($n$ = 8) and social media ($n$ = 10) \add{including Wechat, RED (Xiaohongshu), Reddit, and Facebook}. Participants were invited to complete a screening questionnaire regarding their professions, whether they had used GenAI tools such as ChatGPT and Midjourney before, and which tools they had used. \add{Participants who are not creative professionals (e.g., software engineers) or do not actively use GenAI in their work were not invited to the interview process.}

The interviews were conducted online from April 2023 to July 2023 after receiving institutional review board (IRB) approval. Each interview was scheduled for 60 minutes on video conferencing platforms (e.g., Zoom, Teams, Tencent Meetings) and audio-recorded for transcription. The average duration was 40 minutes, varying from 30 to 60 minutes. Our participants are professionals across different domains, including marketing and PR agencies ($n$ = 3), technology ($n$ = 4), social network ($n$ = 4), cyber security ($n$ = 1), cryptocurrency ($n$ = 1), and non-profit organization (NGO) ($n$ = 1). In addition, our participants have varying job roles, from marketing manager to the founder and chief marketing officer (CMO) of marketing agencies. Our participants are from different countries, including the United States ($n$ = 9), China ($n$ = 8), and South Korea ($n$ = 1). Each participant received a \$20 e-gift card upon completion of the interview. Table~\ref{tab:my_Table} summarizes the characteristics of our participants. 

\begin{table*}{\small
\centering
\renewcommand{\arraystretch}{1.2}
\setlength{\tabcolsep}{2pt}
\caption{\add{Characteristics of participants}}
\label{tab:my_Table}
\resizebox{1.0\textwidth}{!}{
\begin{tabular}{cccccccc}
{\bf Participant} & {\bf Gender} & {\bf Age} & {\bf Job Role} & {\bf Industries} & {\bf Yrs of Exp.} & {\bf Country} & {\bf \add{GenAI Tools}}\\ 
\hline
P1 & Male & 42 & Founder and CMO & PR Agency & 10 & US & ChatGPT \\
P2 & Female &34& Marketing Manager & Technology& 7 & US & ChatGPT, MidJourey, DALL-E\\
P3 & Male & 30 & Marketing Manager & Social Network & 4 & China & ChatGPT\\
P4 & Female & 34 & Senior Marketing Manager & Technology & 5 & US & ChatGPT, Midjourney, Bing\\
P5 & Male & 24 & PR Specialist & PR Agency& 2 & China & ChatGPT, Bard, DALL-E\\
P6 & Male & 40 & Marketing Director & Technology  & 9 & China &ChatGPT, \\
P7 & Male & 23 & Digital Marketing Specialist & Marketing Agency & 2 & China & ChatGPT, Leonardo.AI, Tome\\
P8 & Female & 37 & Marketing Director & Social Network  & 7 & US & ChatGPT, Firefly\\
P9 & Male & 30 & Digital Artist & Creative Advertising  & 7 & China & ChatGPT, Midjourney, Stability.AI\\
P10 & Female & 26 & Data Analyst/Researcher & Legal Services  & 3 & US & ChatGPT\\
P11 & Female & 31 & Marketing Manager & Cyber security & 3 & US & ChatGPT, Midjourney\\
P12 & Male & 25 & Self-media Practitioner & Creative Advertising & 2 & China & ChatGPT, Midjourney, Firefly\\
P13 & Male & 43 & Founder and CEO & PR Agency & 12 & US & ChatGPT\\
P14 & Female & 42 & Founder and CMO & Technology & 10 & China & ChatGPT\\
P15 & Female & 27 & Marketing Specialist & Cryptocurrency  & 2 & South Korea & ChatGPT, Claude\\
P16 & Female & 24 & Digital Communications Manager & NGO & 2 & US & ChatGPT, Microsoft Bing Chat, Bard\\
P17 & Male & 30 & Founder and CMO & Education  & 4 & China & ChatGPT, Midjourney\\
P18 & Male & 34 & Senior Marketing Manager & Social Network & 4 & US & ChatGPT, Midjourney\\
\hline
\end{tabular}}}
\end{table*}

\subsection{Data Analysis}
The dataset for analysis included all 18 interview transcripts. Three researchers (including the first author) manually transcribed the interviews. To ensure transcription accuracy, we carefully examined the data by repeatedly checking back against the original audio recordings. To provide contextual information, each interview began with open-ended questions: {\em i}) Can you tell us about the company/industry you are working in? {\em ii}) Can you tell us your current job responsibilities? {\em iii}) How long have you worked in your current role? This contextual information guided transcribers when coding recordings, especially if they were not the interviewers.

\add{The interviews were largely semi-structured and centered around five major inquires: 
\begin{itemize}
\item What are participants' general perceptions about GenAI? 
\item What are their experiences with specific GenAI tools?
\item How is GenAI integrated into their creative workflows?
\item What complexities and uncertainties do they encounter in this process?
\item How do they develop any strategies to overcome such challenges?
\end{itemize}}
To discover the main themes of the interviews, we followed an inductive approach~\cite{thomas2003inductive} to perform thematic analysis~\cite{braun2006using, braun2019reflecting}. Three trained researchers independently examined transcripts in detail, actively seeking patterns and meanings within the content. Analytic memos were created to record insights. This iterative process ensured a comprehensive understanding of all aspects of the data, aligning with the principles of thematic analysis~\cite{braun2006using}.

Once we gained initial insights into how users implement GenAI in their professional settings, we conducted iterative discussions to refine our understanding, drawing from insights documented in our analytic memos. Employing a constant comparative approach~\cite{glaser2017discovery}, we continually compared emerging categories within the collected data for similarities and differences. Subsequently, we individually assigned basic codes to each idea, with each researcher highlighting and noting relevant text to identify potential patterns. \add{From this coding, we surfaced the main theme about the benefits and challenges around adopting GenAI tools for creative work, which we organized our results around. } Following the guidance of coding `as many potential themes/patterns as possible'~\cite{braun2006using}, we generated a list of \rev{302}{452} basic codes. \add{At the outset of our study, we had some preliminary concepts derived from prior research that examined the nature of creative professionals~\cite{foord2009creativeindustries}. We paid particular attention to situational factors such as the context of using GenAI and the prevailing norms at specific times, based on the situated action theory~\cite{suchman1987situated}.} We held seven one-hour meetings to address coding discrepancies, employing the `Open Discussion' method~\cite{chinh2019ways}. During these meetings, we created a table summarizing the codes applied by each coder to each quote and engaged in discussions to resolve disparities systematically. Coders considered codes applied by others to a given quote and reviewed the reasoning behind each code selection before arriving at a definitive decision. We used codes to facilitate theory development while intentionally avoiding reliance on inter-coder reliability to ensure comprehensive capture of variations and prevent potential marginalization of diverse viewpoints~\cite{mcdonald2019reliability}.

Upon generating the initial coding, we reconvened to compare and discuss the codes and explain how each basic code can be used to represent a potential theme. We then analyzed the codes and decided how different codes could be combined to form a higher-level theme through multiple rounds of discussions. \add{For example, `Comparison among different GenAI tools' and `Developing prompting methods' were combined into a higher-level theme of `Strategies' in the using stage.} After that, we re-examined the candidate themes and refined the themes to ensure internal homogeneity and external heterogeneity~\cite{patton1990qualitative}. Lastly, we defined and named the themes and conducted multiple rounds of refinements before generating the final reports. The final satisfactory thematic map includes three primary themes: \rev{``marvelous,'' ``mysterious,'' and ``maddening'' throughout the ``preparation,'' ``process,'' and ``product'' framework.}{`prospects,' `challenges' and `strategies.'}

\subsection{\add{A Learning-Using-Assessing Framework}}

\add{As illustrated in Fig.~\ref{fig:my_label}, in our study, we examined the human-GenAI co-creation process within a comprehensive LUA framework: \underline{L}earning -- how users acquire GenAI capabilities, encompassing the understanding of concepts, procurement of tools, and development of relevant skills; \underline{U}sing -- how users integrate GenAI into their creative workflows, including defining objectives, exploring possibilities, and refining outcomes; \underline{A}ssessing -- how users evaluate GenAI products, considering multiple aspects such as quality, regulation compliance, and content trustworthiness.} 

\add{Note that while our study is structured into the categories of learning, using, and assessing for clarity of exposition, these activities are often deeply intertwined since human-GenAI interaction often involves a cyclical and iterative process, with various phases overlapping and intersecting. For instance, acquiring new GenAI capabilities and experimenting with them typically occur concurrently, while evaluating intermediate results is essential for the refinement of outcomes. Further, the concrete sequence of activities tends to vary greatly depending on individuals and contexts, reflecting the heterogeneous nature of human-GenAI co-creation in the wild. Therefore, it is worth emphasizing that our framework is not meant to imply a rigid sequence but rather to provide a structured way to understand the different aspects of the human-GenAI co-creation process.}

\section{Findings} 
In our study, we observed that the overall sentiment regarding GenAI \add{in the creative industries} is a blend of optimism and discouragement \add{throughout the process of learning, using, and accessing GenAI, and yet users consistently demonstrate active agency to overcome challenges encountered in the process.} \rev{Thus, we detail our findings describing how participants {\bf prepare} themselves to apply GenAI and the {\bf process} through which they employ GenAI in their workflows, and how they evaluate the final {\bf product} co-created with GenAI. The main findings are summarized in Table~\ref{tab:findings}}{Below, we detail our findings describing 
{\em prospects} -- how participants leverage GenAI to expedite and automate content generation, {\em challenges} -- how they struggle when facing various uncertainties and complexities, and {\em strategies} -- how they actively exercise agency to overcome many of such challenges. Key findings are summarized in Table~\ref{tab:findings}.}

\begin{table*}[t]{\small
\centering
\renewcommand{\arraystretch}{1.2}
\setlength{\tabcolsep}{2pt}
\caption{Summary of key findings}
\label{tab:findings}
\begin{tabular}{clll}
 &\multicolumn{1}{c}{``{\bf Prospect}''} & \multicolumn{1}{c}{``{\bf Challenges}''} & \multicolumn{1}{c}{``{\bf Strategies}''} \\
\hline
\multirow{3}{*}{\minitab[c]{{\bf Learning} \\{ (\msec{sec:learning})}}} & \multirow{3}{*}{\em $\cdot$ Abundant learning resources}  & {\em 
$\cdot$ Scarcity of valuable resources} & {\em $\cdot$ Official documentation first}\\ 
& & {\em $\cdot$ Rapid evolution of GenAI} & {\em $\cdot$ Following influencers' lead} \\
& & {\em $\cdot$ Disparate resource availability} & {\em $\cdot$ Making most of what’s available} \\
\hline
\multirow{4}{*}{\minitab[c]{ {\bf Using} \\ {(\msec{sec:using})}}} & \multirow{4}{*}{\minitab[l]{{\em $\cdot$ Improving efficiency}\\
 {\em $\cdot$ Eliciting human-likeness}\\
{\em $\cdot$ Sparking human creativity} \\
{\em $\cdot$ Transforming creative workflows} \\
}} & {\em $\cdot$ Limited controllability} & \multirow{4}{*}{\minitab[l]{{\em $\cdot$ Selecting the right tools} \\{\em $\cdot$ Developing personalized prompting strategies}}}  \\
 &  & {\em $\cdot$ Ineffective feedback} & \\
  &  & {\em $\cdot$ Engineering-centric design}  & \\
& & {\em $\cdot$ Lack of customizability} & \\
\hline
\multirow{3}{*}{\minitab[c]{ {\bf Assessing} \\ (\msec{sec:product})}} & \multirow{3}{*}{\minitab[l]{{\em $\cdot$ Structured, diversified, and polished} \\{\em $\cdot$ A new form of creativity}}} &
{\em $\cdot$ Authorship disclosure} & \multirow{3}{*}{\minitab[l]{
{\em $\cdot$ Providing proactive transparency} \\
{\em $\cdot$ Situated non-use} \\
{\em $\cdot$ Checking facts manually}}}\\
&  & {\em $\cdot$ Regulatory compliance} & \\
&  & {\em $\cdot$ Content trustworthiness} & \\
\hline
\end{tabular}}
\end{table*}

\subsection{\add{Learning}}
\label{sec:learning}
\add{In this phase, beyond basic concepts and tool usage, users learn how to leverage GenAI's capabilities for their tasks and contexts. This involves exploring advanced features, understanding the nuances of GenAI's responses, and finding users' own methods to achieve the best outcomes.}
\add{For instance, P4, a senior marketing manager, reflected}
\add{\begin{quote}
{\em ``AI-powered marketing has been popular for at least 12 years. However, with GenAI, there are so many new things it could help with. For example, we can now use ChatGPT to write product descriptions. I think the hardest part is to really learn how GenAI works and what its limitations are, and to understand, for example, how it digests our prompts and how we may best leverage its outputs.''} [P4]
\end{quote}
}

\subsubsection{\add{Learning -- Prospects}}
\paragraph{\underline{Abundant learning resources}}
A majority of participants observed a significant increase in online learning resources pertaining to GenAI tools and skills, ranging from courses and tutorials to forums and communities. Some participants highlighted that these resources provide technical knowledge as well as insights into best practices, ethical considerations, and real-world applications of GenAI, which thus not only deepen their understanding of GenAI but also help maintain their competitiveness within their respective fields. For example, P7 noted: 
\begin{quote}
    ``{\em I gather knowledge from platforms such as Bilibili and Red (Xiaohongshu), and some GenAI communities that discuss a range of topics.
    As self-directed learning is key for me, it is great to have access to a variety of resources both in China and overseas.}'' [P7]
\end{quote}
Notably, although these learning resources are becoming more accessible on local media platforms, such as RED (Xiaohongshu) in China [P3, P7, P9], and Naver Band in Korea [P15], almost all participants still identified global platforms like YouTube, Twitter, and TikTok as their main sources for learning. Nevertheless, the highest-rated learning resource among participants is the GenAI courses offered by Deeplearning.AI [P3, P4, P7, P9].

\subsubsection{\add{Learning -- Challenges}}
\paragraph{\underline{\add{Scarcity of valuable resources}}} Despite the wealth of learning resources available, a few participants highlighted the challenge of identifying truly useful information.
\begin{quote}
  ``{\em As ChatGPT has just emerged, I often go to social media to see how other people use it. After several months of trying, I rarely learned from social media or other people about how to use it.}'' [P6]

    ``{\em I think the trial and error is finding the right Youtubers to follow. There is no one-size-fits-all solution, but the tricky part is to find the right people to learn from.}'' [P13] 
\end{quote}
Additionally, there are concerns regarding the \rev{credibility}{quality} of content, as some participants view certain learning resources as merely a means to monetize GenAI products rather than serving educational purposes.
\begin{quote}
``{\em \rev{There are services and books available that sell prompts, provide training, and instruct people on how to use GenAI tools. It's evident that an entire economy is being generated around GenAI.}{Due to the apparent lack of a major centralized online learning platform in China, it seems everyone is seizing the opportunity for profit, which has led to a sense of profiteering. I have personally enrolled in a few paid courses, but the quality varies significantly; some content barely reaches a beginner level.}'' [P3]}
\end{quote}

\paragraph{\underline{Rapid evolution of GenAI}}
Meanwhile, some participants expressed frustration that the GenAI technology changes at an extremely rapid pace, describing it as ``a constantly evolving enigma'' [P8], making it challenging to keep up with its latest developments. As P7 reflected 
\begin{quote}
``{\em I consider myself an expert in adopting GenAI in my field, as I'm always on the lookout for the most recent advancements and make a conscious effort to acquire new skills. Nevertheless, the pace at which GenAI is advancing is astonishing, with each week seeming like a totally new cycle. I sometimes struggle to find the relevant information required to become proficient in it.}''  [P7]
\end{quote}

\paragraph{\underline{\rev{Resource availability}{Disparate resource availability}}}
We also discovered that there is a noticeable difference in the availability of resources for participants in the U.S. compared to those outside the U.S. since accessing US-based resources requires overcoming hurdles such as internet censorship. For example, a participant from China shared: 
\begin{quote}
``{\em I've found it quite challenging to keep up with the latest GenAI tutorials on social media since many of them are blocked in my country.}'' [P17] 
\end{quote}
\add{In addition, participants who are non-native English speakers reflected the challenges due to lack of learning resources in their languages:} 
\begin{quote}
``{\em My proficiency in English is decent, but technical jargon can be overwhelming. I wish there were more resources in Chinese, which would make learning much smoother.}'' [P17] 
\end{quote}

\begin{quote}
``{\em I've found it quite challenging to keep up with the latest GenAI tutorials on social media since many of them are blocked in my country.}'' [P12] 
\end{quote}

\subsubsection{\add{Learning -- Strategies}}
\paragraph{\underline{\add{Official documentation first}}}
\add{Participants [P1, P3, P4, P6, P14, P17] reported that they primarily depend on the official documentation of GenAI tools for learning, despite the challenging nature of self-directed learning.}
\begin{quote}
     \add{``{\em I think the primary learning resource would be the official documentation, of which the information is more credible, even though it means that I have to learn the materials by myself.}'' [P2]}
\end{quote}
\add{This observation intriguingly indicates that participants appear to be balancing the credibility of resources against the ease of learning, with a tendency to prioritize the former over the latter.}

\paragraph{\underline{\add{Following influencers' lead}}}
\add{Further, participants [P1, P5, P7, P11, P12, P14, P17] indicated that social media remains the most effective channel for acquiring new information, observing advancements, and comprehending emerging trends in the GenAI field.}
\begin{quote}
    \add{``{\em I typically visit Twitter to keep up with the latest trends and check for updates on GenAI tools. Many YouTubers share their experiences of using GenAI tools to create original content. I find their work very creative and inspiring, which has certainly influenced my own creation.}'' [P6]}
\end{quote}
\add{The quote also highlights the significant role of content creators, such as YouTubers, who experiment with GenAI tools and share insights into their creative processes.}

\paragraph{\underline{\add{Making most of what's available}}}
\add{China-based participants shared their strategies to overcome the access issue due to the Great Firewall.}
\begin{quote}
    \add{``{\em To navigate the restrictions of the Great Firewall, we've had to get creative. We often use VPNs and rely on peer-to-peer networks to share necessary tools and resources for our work}'' [P12]}
\end{quote}
\add{For participants lacking technical backgrounds with VPN, a common strategy appears to be maximizing the utility of the limited learning resources available. For example, P3 shared his approach to learning:
}
\begin{quote}
     \add{``{\em Andrew Ng's courses on Deeplearning.AI are quite popular in China because they are free and accessible. To make the most of these courses, I first practice ChatGPT by myself and then compare my work with the teachings in the courses to find areas for improvement.}'' [P3]}
\end{quote}

\subsection{\add{Using}}
\label{sec:using}
\add{In this phase, users integrate GenAI into their daily tasks or creative workflows, which involves defining objectives, selecting tools, exploring possibilities, and refining outcomes.}


\subsubsection{\add{Using -- Prospects}}

\paragraph{\underline{\add{Improving efficiency}}}
\rev{Previous research has examined how practitioners in the creative industries may leverage AI's analytical power to automate repetitive functions and activities~\cite{huang2021strategic,shane2004planning}. Some participants also consider GenAI tools to provide great opportunities to enable professionals to achieve their objectives more independently, which potentially changes the traditional workflow in the creative industry.}{One prominent benefit of GenAI identified by our participants is its impact on the efficiency of content creation. Participants consistently highlighted how GenAI revolutionizes the content creation processes by significantly reducing its required time and labor.}
\begin{quote}
``{\em For the advertising industry, before the advent of GenAI, creating an image like this (one created by Midjourney) would involve 3D rendering and post-production editing, which could be exceedingly expensive. So the biggest advantage of GenAI is reducing costs and increasing efficiency.}'' [P9]
\end{quote}

\paragraph{\underline{\add{Eliciting `human-likeness'}}}
GenAI provides an open-ended, conversation-based interface that encourages exploration and experimentation. \add{Our participants reflected such affordance as `a natural way of communicating ideas with a machine' [P3] and found such interfaces `easy to explore the outputs through trials and errors' [P9], which also elicit the perceived `human-likeness' of GenAI.} 
\begin{quote}
    {\em ``I really appreciate the chatty interface of ChatGPT. It often feels like I'm having a conversation with my other human colleagues, which is quite remarkable. The way it engages in dialogue and provides responses in a natural, conversational manner is impressive.''} [P8]
\end{quote}

\paragraph{\underline{Sparking human creativity}}
\add{Our participants observed that GenAI can be especially beneficial during the ideation phase.} Participants [P1, P3, P4, P6, P17, P18] shared their experience of exploring GenAI as valuable resources to generate initial ideas as `starting points':
\begin{quote}
``{\em When we use ChatGPT to map out the marketing strategy, user acquisition approach, and advertising expenditure, it proves quite effective to lay down the initial rough draft.}'' [P3]
\end{quote}
In addition, participants [P7, P12, P18] also mentioned using GenAI for brainstorming: 
\begin{quote}
    ``{\em I use GenAI to assist in generating content. For example, Tome can be helpful in the early stages of creating stories or script concepts.}'' [P7]
\end{quote}
Other participants noted that GenAI may occasionally produce `surprising' results that spark their creativity and enable them to tackle tasks with new and inventive perspectives. For instance, P17 recounted his experience of using ChatGPT to devise marketing strategies:
\begin{quote}
    ``{\em Once you feed ChatGPT with your initial directives, it can produce a multitude of ideas, which you can expand on infinitely or continually delve deeper. You may find that it uncovers insights you have not considered before. This can be quite surprising.}'' [P17]
\end{quote}

\paragraph{\underline{\add{Transforming creative workflows}}}
\add{Interestingly, as GenAI becomes integral in the ideation phases, it also has the potential to transform traditional creative workflows profoundly.} \rev{who has been employing GenAI tools to craft marketing campaigns, has transitioned to being a full-time creator of artificial intelligence-generated content (AIGC)}{For example, P9, who previously relied on a team for executing marketing campaigns, has now shifted to using various GenAI tools. This change has notably expanded his capacity to handle a wide range of tasks independently, enabling him to transition into a self-employed, independent digital artist}:
\begin{quote}
    \add{``{\em My current role is different from a traditional advertising agency; I'm creating artistic work, which is more about generating, exploring, and combining various AI-generated content (AIGC). }'' [P9]}
\end{quote}
\add{P9 further explained that GenAI affords him increased autonomy in his work, enabling him to explore artistic freedoms that would have been unattainable in traditional settings:}
\begin{quote}
\add{"{\em I made it clear to my clients from the outset that our relationship is one of `co-creation': I'm not designing for you in the traditional sense, so I can't fulfill your specific design requirements. Instead, I need a theme, a broad concept, and ample room to train AI to generate something intriguing.}" [P9]}
\end{quote}
\add{As P9 observed, GenAI transcends being merely a tool and becomes a `partner' or a `collaborator' in the human-GenAI co-creation process. With GenAI's help, creative professionals are now afforded to move beyond traditional roles and explore more artistic and creative freedoms.}
\add{Yet, it is worth noting that while some participants in the marketing industry 
share P9's perspective on the disruptive role of GenAI, others view it merely as an additional tool, not markedly different from previous AI-powered tools used in marketing. For instance, P4 highlighted this viewpoint:}
\begin{quote}
\add{``{\em AI is not a new concept. Data-based marketing has been trending since the 1980s, so it doesn't just build up all of a sudden. Now with Gen-AI, we still need to do similar business.}'' [P4]}
\end{quote}



\subsubsection{\add{Using -- Challenges}} \hfill

\vspace{8pt}
\add{Despite the aforementioned optimism,} participants also admitted that GenAI is in its early stage and its adoption in creative industries is still `limited' and `unsystematic' \add{[P3]}. As P13 summarized: 
\begin{quote}
``{\em GenAI is incredibly promising, yet it's still in its very early days. It's not the panacea that many think it is. In fact, the more one engages with it, the more apparent its limitations become.}'' [P13]
\end{quote}
\add{Specifically, participants reflected on major challenges that hinder the effective use of GenAI in the creative industries: limited controllability, ineffective feedback, engineering-centric design, and lack of customizability.}




\paragraph{\underline{\rev{Prompt, prompt, prompt}{Limited controllability}}} 
 A recurring concern among participants [P1, P3, P7, P8, P9, P13, P18] is the lack of controllability in existing GenAI tools compared with more conventional tools such as Photoshop. \add{Achieving desired outcomes often involves iterative trials, due to the highly non-deterministic, unpredictable nature of GenAI's generative process.} 
\begin{quote}
    ``{\em When I use Midjourney, I feel like the more I want to control the outcomes, the less likely it can meet my needs. It's still in the early stage and lacks user-friendly features like Photoshop, which can easily translate your ideas into reality. GenAI introduces a higher degree of unpredictability and you have to fine-tune through extensive testing and, well, a bit of luck!}'' [P8]
\end{quote}
Participants, especially the novices of GenAI tools [P1, P2, P5, P10, P12, P15, P17], expressed \rev{confusion about how to obtain desired outcomes through prompting}{their frustrations when these tools fail to produce desired results, even after many rounds of experimentation.}
\begin{quote}
``{\em It's challenging to guide GenAI tools to produce desired results. For instance, I tried to use GenAI to create a cover page for our website, showing a road split into two directions towards the horizon. I probably tried 25 different prompts, such as `a single road split into two roads that go east and west towards the horizon,' but neither Midjourney nor Leonardo.AI were able to generate the results I was looking for.}'' [P13] 
\end{quote}
\add{In a similar vein, P2 expressed her uncertainty.}
\begin{quote}
\add{``{\em Sometimes it's extremely difficult to get the results right. Midjourney kinda gets what I think, but there is always something missing. It often takes many iterations to get there. I feel there might be issues with how I'm using it or the limitations of current technology. I'm not sure.}'' [P2]} 
\end{quote}
\rev{Further, employing GenAI to craft intricate concepts often requires not just choosing the right tools but also finding exquisite ways to maneuver those tools.}{Further, participants reflected that it becomes even more challenging when the desired outcomes involve creative and complex concepts.} P9 shared his experience \rev{of using Stability.AI's Stable Diffusion to}{of using Midjourney to} produce advertising art (see Fig.~\ref{fig:sample.png}). 
\begin{quote}
    ``{\em When working with a food delivery brand, I had this complex concept: a boat floating on the Suzhou River pier at the Bund in Shanghai, with the boat being shaped like a wave, carrying a cup of takeaway coffee.  Midjourney can't understand such complex concepts; instead, it created either a coffee cup floating on the river surface or a ship next to the Bund dock, but it can't transform the ship into the shape of waves, nor can it place the coffee on top of the waves, because this idea is too complex.}'' [P12]
\end{quote} 
\add{Thus, GenAI's capacity to generate a virtually limitless array of outcomes also makes it difficult to precisely control and generate content that aligns with users' expectations and visions.}

\paragraph{\underline{\rev{Feedback or not?}{Ineffective feedback}}} \add{Our participants also reported another relevant issue with current GenAI tools is the absence of effective feedback channels. When users face technical challenges or less-than-ideal outcomes, there are no straightforward ways to suggest enhancements or modifications. They also highlighted that the design of existing feedback mechanisms, such as the thumb-up/-down option in ChatGPT, is surprisingly ineffective or non-intuitive.}
\begin{quote}
   \add{``{\em It's similar to asking a person but not receiving the response you hope for. Instead of pressing a button to show `like' or `dislike', you would provide more clarification or explain your question better. This would help the person understand and respond more accurately to your question, much like a human-to-human conversation.}'' [P16]}
\end{quote}
\add{Additionally, P4 noted that while a thumbs-up/thumbs-down feedback function might assist engineers in system tuning, it fails to provide users with the `immediate, real-time improvement' [P4] they seek in conversational interaction. This highlights a gap between user expectations and the feedback mechanisms in current GenAI tools.}

\paragraph{\underline{Engineering-centric design}} Another concern echoed by participants is the engineering-centric design of some GenAI tools, which poses a challenge for non-technical users. For instance, P9 recounted his struggles in using GenAI tools for animation creation:
\begin{quote}
    ``{\em Crafting animations with GenAI can be difficult, as many of these tools are designed by programmers rather than artists, making their operational logic hard for me to understand. For example, to create a scene, I must define parameters like various axes and coordinates to control camera movements. Often, this involves manual calculations or even programming in Python to determine these parameters. As a designer by training, I find this aspect particularly challenging. I wish there were more intuitive interfaces, like a PlayStation controller that would allow me to control the camera movements more naturally.}'' [P9]
\end{quote}

\begin{figure}
  \centering
    \includegraphics[width=0.36\textwidth]{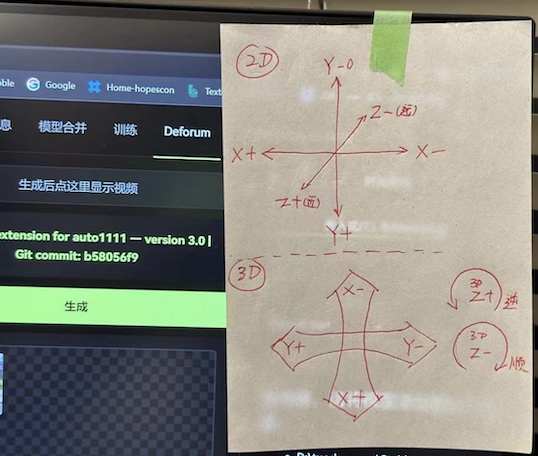}
\caption{P9's `cheat-sheet' when using Deforum Stable Diffusion for animation creation. \label{fig.handwriting.png}}
\end{figure}

As shown in Fig.~\ref{fig.handwriting.png},\footnote{The permission to include this image in the paper has been officially granted by P9.} P9 even shared his handwritten `cheatsheet' with us, highlighting this demanding process.

\paragraph{\underline{\rev{Customized generation}{Lack of customizability}}} 
Participants [P1, P2, P9] also shared their struggles to use existing GenAI tools to generate localized content, which is often required for their domain-specific tasks. For instance, P2 recounted:
\begin{quote}
    ``{\em I notice that ChatGPT often produces quite generic information. When asked about highly specialized subjects, it still tends to reply in a generic manner. I understand this might be due to the nature of its training data. I believe if it had access to more domain-specific training data, it would be more appropriate for specialists in various fields and easier to use.}'' [P2]
\end{quote}
As a China-based digital artist, P9 remarked:
\begin{quote}
    ``{\em In my creation process, I often face the challenge of depicting local elements. This is mainly because the training data for these tools contains little domestic content, making it challenging to create local landmarks such as Shanghai's Oriental Pearl Radio and Television Tower.}'' [P9]
\end{quote}
\add{Also, GenAI tools often struggle to accurately capture or portray non-western cultural contexts in their outcomes. For instance, P17, who is the founder of an education start-up in China, noted:}
\begin{quote}
    \add{``{\em One of our projects involved creating AI-illustrated children's books. We wanted to depict the proverb of `grinding an iron rod into a needle,' a story about Li Bai, one of the greatest poets of the Tang dynasty. However, Midjourney couldn't grasp the context, and we had to describe the story in detail in English, which inevitably lost certain meanings.'' [P17]}}
\end{quote}

\subsubsection{\add{Using -- Strategies}}\hfill

\vspace{8pt} 
\add{We discovered that rather than passively awaiting for advancements in GenAI tools, participants proactively exerted their user agency to overcome some of the above challenges, especially the limited controllability of GenAI tools.}

\begin{figure*}[ht!]
\centering
\includegraphics[width=0.7\textwidth]{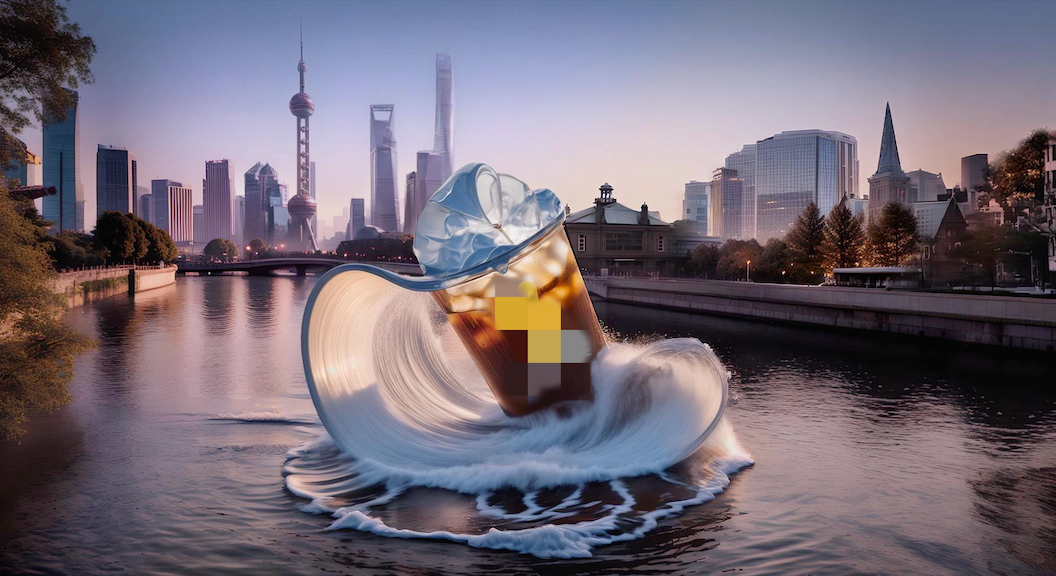}
\caption{Sample advertising art created by P9 using Stability.AI's Stable Diffusion, \add{which represents the complex concept of `a boat floating on the Suzhou River
pier at the Bund in Shanghai, with the boat being shaped like a wave, carrying a cup of takeaway coffee.'}}
\label{fig:sample.png}
\end{figure*}

\paragraph{\underline{\add{Selecting the right tools}}}

\add{To overcome the controllability challenge, several participants [P1, P3, P7, P9, P13, P18] concurred on the importance of finding the appropriate GenAI tools. For example, P9 shared his experiences with text-to-image generative tools and encountered difficulties in controlling particular elements within the generated images. After numerous searches and trials, he eventually switched to a specific tool that offers the requisite control:}
%
\begin{quote}
    \add{``{\em After an extended period of unsuccessful trials with Midjourney, I eventually found Stability.AI's Stable Diffusion, which provides hierarchical control that meets my requirements. I found a delicate approach to realize my concept: I rendered the background and foreground separately; after importing some modelings of the Bund, I used the control network to reconstruct the silhouette and details of the Bund's buildings with about 80\% to 90\% accuracy. As you can see (Fig.~\ref{fig:sample.png}\footnote{The permission to include this image (with the brand logo omitted) in the paper has been officially granted by P9.}), the background in the generated image appears highly realistic.}'' [P9]} 
\end{quote}
\add{In addition, participants also recognized the limitations of current GenAI tools and adopted a hybrid approach by leveraging the strengths of both GenAI and conventional tools. As P12 shared:}
\begin{quote}
    \add{``{\em When creating commercial posters for a fashion company, we use Midjourney to generate the model's image and background. However, Midjourney doesn't produce a perfect image in a single attempt, so I have to deconstruct the requirements. I first extract the figure to create a transparent background. Then, I generate the corresponding background, ensuring the right angle and lighting atmosphere. After that, I use Photoshop to assemble these elements, ultimately crafting the desired image.}'' [P12]} 
\end{quote}
\add{P12 further explained his rationale:}
\begin{quote}
    \add{``{\em Midjourney is better for generating a wide range of high-quality, photorealistic images; Photoshop is better for its superior image-blending capability, which allows for precise adjustments and seamless integration of details within an image's background. In our workflow, it's essential to harness the best features of each software to achieve control over the final imagery.''} [P12]}
 \end{quote}
\add{P12's practice reflects a deep understanding of the capabilities and limitations of each tool, demonstrating the importance of a nuanced approach to integrate GenAI into creative workflows.}

   

\paragraph{\underline{\add{Developing personalized prompting strategies}}}
\add{Moreover, while participants shared their frustrations due to the limited control over GenAI's processes and outcomes,} through trial-and-error and mental models, a few participants developed their own prompting tricks. For example, P12 noted: 
\begin{quote}
    ``{\em It requires a lot of practice. Once you become adept, you begin to get a handle on which words might work. For example, there is a particular word to describe `whole body' in your vocabulary. While there are many other options, you stick to this specific word as it offers a high level of accuracy.}'' [P12]
\end{quote}
As a digital artist, P9 used text-to-image GenAI tools more often and shared his strategy as a `one-sentence' guideline: 
\begin{quote}
    ``{\em My approach is to guide GenAI by articulating my ideas in a single sentence. This approach has also shaped my prompting method. Each image is created to express just one idea, with subsequent descriptive terms serving merely as guides for time, environment, lighting, and other specialized aspects of the content.}'' [P9]
\end{quote}
\add{We also noticed that participants tend to develop their prompting strategies based on their own understanding, or `mental models'. For example, P12 shared:}
\begin{quote}
    ``{\em From my experience, I observed that ChatGPT seems to place more emphasis on words closer to the beginning of the prompt. If the result does not convey a specific meaning I intend, I assess whether that word is positioned too far back in the prompt and make adjustments step by step.}'' [P12]
\end{quote}
\add{A mental model essentially describes how a user perceives, understands, and predicts how a system works~\cite{allen1997mental}. In this context, P12 developed his own mental model of how word positioning affects GenAI's output, leading to a specific prompting strategy.} 
For participants with greater expertise, self-guided learning, and trial-and-error helped them develop their own systematic strategies for prompting. For instance, P3 outlined four key steps--role assignment, task contextualization, task decomposition, and task specification--as part of his approach to refining ChatGPT's output:
\begin{quote}
    ``{\em Essentially, you interact with ChatGPT through conversations, and conversations inherently involve roles and tasks. Therefore, in my view, the first step is defining the role. For example, if you tell it that you are a journalist, it will narrow down its knowledge base to align with the journalist's context. Next, you provide it with specific tasks, the background of the tasks, their requirements, and objectives. Then, it involves breaking down the tasks, ensuring each task has a final assessment or feedback, and presenting it accordingly.}'' [P3]
\end{quote}


\subsection{\rev{Product}{Accessing}}
\label{sec:product}

\add{In the assessing phase, users evaluate the content generated by GenAI  from multiple aspects including quality, creativity, regulation compliance, and  trustworthiness.}

\subsubsection{\rev{Marvelous}{Assessing -- Prospects}}

\paragraph{\underline{Structured, diversified, and polished}}
Participants were often impressed by the quality of the content generated by GenAI. Especially in language-related tasks, they found that GenAI excels at structuring ideas, producing a variety of alternatives, and refining written work. For example, P4 remarked:
\begin{quote}
    ``{\em I have drafted some editorials for our hypercasual game apps. I use ChatGPT to generate variations of the text. I ask, `Could you generate 10 different versions of the text within these constraints and suggest titles I can use?' Usually, I get quite pleasing results, which I then modify into messages for our customers.}'' [P4]
\end{quote}
For non-native English speakers, GenAI tools such as ChatGPT significantly enhance the quality of their writing. For example, both P2 and P11 observed:
\begin{quote}
    ``{\em I often use ChatGPT to refine my writing to a specific tone, whether it is more direct or formal. As non-native speakers, we can depend on GenAI to improve the language. Sometimes, we even start by drafting in Chinese because it is quicker, then use GenAI to translate it into English. We then make minor tweaks and revisions to get the final version.}'' [P2]
\end{quote}
\begin{quote}
    ``{\em It seems that ChatGPT understands what I am looking for. The quality of its writing is better than what I could write myself.}'' [P11]
\end{quote}
\paragraph{\underline{\add{A new form of creativity}}}
Further, a few participants reflected that, due to its vast computational power and memory capacity, GenAI possesses a more extensive `knowledge base' [P2, P3, P11] and a broader `search space' [P3] compared to humans, fostering `a new form of creativity'.
\begin{quote}
    ``{\em People can perceive this vast gap between our knowledge, which may cover from 1 to 100, and ChatGPT's potential knowledge, which may span from 1 to 100 billion. Anything beyond 100 may seem like a display of creativity to us. Creativity, in practice, does not always need entirely original ideas but involves leveraging existing knowledge, reorganizing it, and generating new content.}'' [P3] 
\end{quote}
\add{Interestingly, several participants further embraced the inherent unpredictability of GenAI's outcomes as one contributing factor for its creativity.}
\begin{quote}
    ``{\em Using GenAI is like opening Pandora's box; there is high uncertainty in the responses it will generate. Often, one must adopt a very open and optimistic mindset, allowing it to create.}'' [P9]
\end{quote}
\add{In comparison, several participants held more nuanced views about GenAI's creativity. For instance, P7 elaborated:}
\begin{quote}
    ``{\em If one asserts that GenAI has creativity, I contend that it actually depends on human input. However, if one argues that it lacks creativity, we might be overlooking the vast amount of data underpinning it. I believe the volume of this big data goes beyond what humans can comprehend or manipulate, offering new and interesting possibilities.}'' [P7]
\end{quote}
\add{Thus, P7 was of the opinion that while GenAI's creativity is largely shaped or influenced by the data and instructions provided by humans, the extensive volume of training data it possesses opens up numerous new possibilities, enabling it to generate unique and novel outputs.}
Meanwhile, other participants [P1, P2, P4, P12, P14] disputed the notion that GenAI possesses innate creativity, arguing instead that the outcomes are simply a manifestation of the users' creativity, with GenAI tools merely assisting users in articulating their creative ideas: 
\begin{quote}
    ``{\em If you want good results, you need to guide Midjourney well, but whether the outputs are creative really depends on your own creativity: If you are a beginner, you may be impressed by its outputs very easily and think, oh, this is surprisingly creative; if you are more experienced and professional, the results will be professional-level creative. It is strongly correlated with your expertise.}'' [P12]
\end{quote}
\subsubsection{\rev{Mysterious}{Assessing -- Challenges}}
\paragraph{\underline{Authorship disclosure}} When inquired about their views on authorship, most participants [P1, P3, P4, P7, P11, P12, P17] perceived the final product as a combination of human creativity and GenAI augmentation, resulting from a fascinating human-GenAI collaboration, in which it is difficult to identify the contributing origins of particular elements. Therefore, participants were divided about the necessity of disclosing whether a product was created by GenAI, based on their perception of GenAI tools, consumer expectations, and legal regulations.
For example, some participants considered GenAI tools as merely one of many resources that assist them in achieving their goals, and therefore, they adhered to the same guidelines as when using other tools. For example, P1, a founder of a PR agency, remarked:
\begin{quote}
\emph{``It's like Google or Bing. There is no need to tell your clients that you get that information on Google or Bing. There is no need to disclose the research tool you have used. It is all about the content quality.''} [P11]
\end{quote}
In addition, participants [P1, P7, P11, P12] deliberately refrained from openly admitting their use of content generated by ChatGPT, citing perceived violations of social norms. This sentiment was echoed by their colleagues, who also chose not to disclose their use of such content. For example,
\begin{quote}
\emph{``It's probably not good to say `I created this using ChatGPT'. My coworkers also do not disclose any such information.''} [P11]
\end{quote}
Another factor contributing to the hesitancy in revealing authorship is the anticipated expectations of consumers. Participants [P1, P3, P4, P11, P17] shared a common belief that customers do not expect to receive such information, and quality matters more than the author of the content:
\begin{quote}
   \emph{``I don't think the reader is going to ask whether there is GenAI behind it. The reader just wants to read this article, and if it is a terrible article, it doesn't matter whether it is written by a human or GenAI.''} [P1]
\end{quote}
Meanwhile, some participants underscored the importance of acknowledging and tagging the content co-created with GenAI, indicating that this practice is essential either to adhere to platform regulations, avert potential copyright legal challenges, or address ethical dilemmas:
\begin{quote}
    ``{\em If it is the content that the client intends to use externally, it's imperative to determine the ownership of the copyright in advance because brands are undoubtedly concerned about who holds the copyright and whether it can be distributed through public channels.}'' [P7]
\end{quote}

\paragraph{\underline{\rev{Evolving regulation}{Regulatory compliance}}} Several participants also highlighted the complex and evolving regulatory environment surrounding GenAI, which influences the way businesses handle the use of content generated by GenAI tools, particularly in contexts such as international marketing. For example, P4 remarked:
\begin{quote}
    ``{\em When we launch marketing campaigns, our target regions typically include the US, Canada, Europe, and Japan. Japan recently introduced laws allowing the use of images generated by GenAI. However, the G-7 countries and the US are currently in negotiations for stricter regulations. This is an aspect we need to be cautious about.}'' [P4]
\end{quote}

Additionally, many companies have implemented a confidentiality compliance framework for the use of GenAI, which forbids inputting sensitive organizational or personal information into public GenAI tools like ChatGPT. This compliance may also significantly limit the utilization of GenAI tools, as P4 shared:
\begin{quote}
    ``{\em In our company, we use a variety of GenAI tools that don't process any data from the company. The customer-facing documents we request them to generate are usually very generic. If we try to use GenAI to process our own articles, the results may contain highly confidential information that is prohibited from being shared externally.}'' [P4]
\end{quote}

\paragraph{\underline{Content trustworthiness}} Moreover, participants [P1, P13, P16] experienced a significant decline in trust towards GenAI-generated content when they encountered inaccurate information (or hallucinations), especially when drafting factual documents such as press releases.

Several participants [P7, P13] noted that at times, ChatGPT fails to adhere to specific restrictions and inadvertently includes forbidden words or expressions despite being instructed otherwise, which considerably undermines trust in GenAI-created content. As P13 reflected:
\begin{quote}
``{\em In a particular project, we had a lot of legal restrictions regarding what we could say or how we could present a product in a blog. Therefore, I instructed ChatGPT to steer clear of specific terms. However, upon reviewing the text, I noticed that it still included variations of the prohibited words. Had we not checked the context, we could have faced serious issues. }'' [P13]
\end{quote}

\subsubsection{Assessing -- Strategies}
\paragraph{\underline{\add{Providing proactive transparency}}}
\add{Several participants conveyed that, in the absence of clear guidelines for content creators regarding the disclosure of GenAI usage, they prefer to play it safe by being transparent about GenAI's role in their content. This approach is aimed at avoiding potential copyright issues. As P8 observed:}
\begin{quote}
    \add{``{\em If written by ChatGPT, we won't label the content as a PR copy but rather an advertisement. We'll include a note stating `written by AI', along with a disclaimer. To avoid copyright issues or legal disputes, we're exercising extra caution in our use of ChatGPT.}'' [P8]}
\end{quote}
\add{P8 stressed differentiating between PR material and advertising, advocating for clear labeling of GenAI-generated content. This practice can be considered as a proactive transparency strategy to avoid potential copyright issues.}

\paragraph{\underline{\add{Situated non-use}}}
\add{In response to issues of a compliance violation, several participants ceased their use of GenAI accordingly. For example, after discovering that ChatGPT was unable to consistently avoid certain terms, P13 decided to \emph {`stop using ChatGPT in similar situations that could potentially result in severe consequences'} and instructed his team to adopt the same approach.} \add{P4  noted that in marketing cases involving countries with unclear or differing regulations on GenAI, their company advised to \emph{`avoid using GenAI when collaborating with them.'} This `non-use' strategy~\cite{baumer2015importance} reflects a cautious approach towards the utilization of GenAI in contexts where compliance is a sensitive issue.}

\paragraph{\underline{\add{Checking facts manually}}}
To mitigate the challenge of content trustworthiness, some participants resorted to cross-checking to verify information accuracy, while others opted to refrain from using GenAI in situations where content factuality is crucial.
\begin{quote}
    ``{\em I do cross-checking because ChatGPT occasionally fabricates information randomly. It's crucial to verify the information, especially when I'm writing a PR press release that requires accurate details; otherwise, it could lead to big troubles.}'' [P12]
\end{quote}
A few participants [P6, P13, P15] believed that existing GenAI tools inherently require human intervention for comprehensive understanding and contextual adaptation. For example, P15 noted:
\begin{quote}
``{\em ChatGPT and Bard change the way we do searches. However, you still need to do your due diligence. For example, if you use ChatGPT to draft a blog post or a press release, there is no `one-click' solution. You always need to do manual verification and curation to make sure the information is accurate.}'' [P15]
\end{quote}

\section{Discussion}

\rev{Our study unveils users' diverse strategies to effectively accomplish their goals and tasks using GenAI. Users describe a spectrum of experiences when integrating GenAI into their daily work, perceiving GenAI tools as a combination of marvelous, mysterious, and occasionally frustrating elements. While existing research in HCI has focused on how GenAI supports specific tasks or develops technical mechanisms to enhance its functionality, our study expands upon these efforts by showcasing the highly varied ways in which GenAI is integrated into creative workflows. These workflows encompass a range of tasks, including ideation, iteration, validation, and evaluation. In addition, our study discovers users' learning of GenAI as a new challenge associated with GenAI development, which has yet to be discussed in the current discourse of GenAI in HCI. We also contribute to the recent discussion about ethical concerns over GenAI by providing empirical evidence of how users perceive and practice ethical decisions in real-world settings.}{Our study made multiple contributions to the existing HCI literature on learning and adopting emerging AI technologies. {\em i}) We uncovered nuanced views about GenAI's unique attributes, especially its creativity, within creative industries, where uniqueness and novelty of content are critical; {\em ii}) Through the lens of a holistic learning, using and assessing framework, we illustrated that users interact with GenAI through a dynamic, evolving human-GenAI co-creation process; {\em iii}) Beyond GenAI's functional limitations, we identified many new challenges (e.g., authorship disclosure) faced by users `in the wild,' showing how users perceive and navigate such decisions in real-world contexts. {\em iv}) We revealed a proactive engagement among users who exercise agency to overcome GenAI's limitations and strategically utilize its strengths to enrich the creative process.}

\subsection{\add{Perceptions of GenAI in Creative Industries}} 
\subsubsection{\add{Perceived GenAI's creativity}}
\add{Reflecting on recent research by Gero et al.~\cite{gero2023social} about creative writers' views on GenAI's creativity, our study found that users in creative industries hold mixed opinions. Some recognize GenAI's vast computing power and memory capacity, considering it a new form of creativity, while others deem GenAI-generated content as `mediocre', attributing its creation to the mere replication of existing data and embracing the unpredictability of GenAI's outputs ~\cite{ganguli2022predictability}. Further, our study showed that perceptions of GenAI's creativity vary with task requirements and users' expertise with GenAI tools. For example, realizing intricate concepts that involve multiple abstract layers (as in P9's case) places a higher demand on GenAI compared to simpler tasks, such as generating various versions of product descriptions (as in P3's case), significantly impacting how GenAI's creativity is perceived. Additionally, expert users may be more critical of GenAI-generated content than novice users. Our findings highlight the intricate, multi-dimensional views on GenAI's creativity, suggesting further research into defining GenAI's creative capacity, including insights from the broader perspective of machine creativity~\cite{jordanous2009evaluating}.}

\subsubsection{\add{Perceived GenAI's role}}
\add{We also observed a split in how participants perceive GenAI's role. While some regard it simply as a `tool' for enhancing efficiency, others believe it represents a transformative force capable of revolutionizing traditional practices. In the latter case, users see GenAI as a `partner,' leading to new forms of human-AI collaborations. Our study extends previous literature on how users' expectations and evaluations of new AI technologies are shaped by their perceptions of such technologies~\cite{chiang2020exploring, gursoy2019consumers}. This perception variation is linked to the degree to which users anthropomorphize GenAI, attributing human-like qualities and abilities to it. While our findings resonate with recent studies~\cite{ma2023users} that perceived humanness greatly influences GenAI usage, we also revealed that viewing GenAI as a partner, rather than a tool, primarily arises from users' assessment and recognition of `machine agency'~\cite{sundar2020rise}. This divergence in user perspectives highlights GenAI's potential to redefine AI's roles and functions in creative industries, promoting a paradigm shift from mere tool usage to creative collaborations to enhance human creativity~\cite{HBR2023GenerativeAI}.}

\subsection{\add{Uses of GenAI in Creative Workflows}}
\add{Integrating GenAI into creative workflows represents a continuous, iterative process that involves learning, experimenting, and assessing. For example, P9, a digital artist, acquired GenAI skills by engaging with social media and experimenting with various prompting techniques. After developing a solid understanding of text-to-image generation, he expanded into video generation using GenAI. Yet, this progression also presented new challenges (e.g., engineering-centric interfaces of text-to-video generative tools). These experiences highlight the evolving nature of adapting GenAI in creative workflows, where switching to each new context tends to require new learning and adaptation strategies, as suggested in the Situated Action Theory~\cite{suchman1987situated}.}

\add{Our study enriches existing HCI literature by exploring human-(Gen)AI interaction from the perspective of a dynamic, evolving process rather than a static, one-time instance. Through the lens of longitudinal learning, using and assessing framework, our study sheds light on the gradual process of how users acquire new GenAI skills, the significance of community-based learning, and the iterative nature of GenAI adaptation in creative workflows~\cite{rafner2023creativity}. This approach offers a more thorough understanding of how human-GenAI interaction evolves over time. Further, it provides detailed and nuanced insights into how GenAI influences each stage of creative workflows and labors~\cite{frank2019framework}.}

\subsection{\rev{Ethical Concerns}{Non-functional Challenges of GenAI}}
Besides GenAI's functional limitations, our study revealed a multitude of challenges encountered by real-world users. For example, the authorship of work co-created by humans and GenAI is distinctive from that created solely by GenAI~\cite{rezwana2023user}. \add{Thus, when utilizing GenAI for brand creation, concerns emerge regarding authorship and copyright infringement~\cite{epstein2023art}.} Our study revealed the intricate ways in which practitioners individually confronted and navigated complex ethical challenges, especially when GenAI was operated in a `grey area' lacking clear policies or regulations~\cite{Greay, hacker2023regulating}. Participants were divided in their views on the necessity of disclosing GenAI usage. Some believed that disclosure is crucial for mitigating potential copyright risks, while others compared using GenAI to a standard Google search, seeing no need for disclosure. Also, the absence of clear guidelines on data privacy protection impeded GenAI's adoption within enterprises~\cite{sun2023automl}. Therefore, our study contributes to the ongoing discourse of developing responsible and ethical GenAI~\cite{muller2023genaichi23}, in which it is essential to consider both individual concerns and contextual factors, such as its use within corporate and enterprise settings.

\subsection{User Agency with GenAI in the Wild}
\subsubsection{User agency in selecting right tools}
Activity Theory centers around how users adapt and use tools to enhance their abilities to attain specific goals~\cite{bodker1989activity}. According to this theory, users `appropriate' tools to facilitate their goals and often use a combination of tools rather than relying on a single one~\cite{suchman1987situated}. This recognition highlights the complexity of many tasks, where a variety of tools may be necessary for users to attain their desired outcomes. Our study uncovered that users selectively apply GenAI for ideation and organization, but much less so in contexts where factual accuracy is critical. \add{Moreover, users combine GenAI's capabilities with traditional tools like Photoshop to produce high-quality visuals,  recognizing the strengths and limitations of each tool [P12].} These findings resonate with previous research~\cite{sun2023automl}, where users actively select different AutoML tools and platforms after careful evaluation. In essence, our study highlights how users creatively employ multiple tools to empower themselves in various contexts. 

\subsubsection{User agency in \rev{developing prompting methods}{refining GenAI's outcomes}}
GenAI is inherently non-deterministic and unpredictable~\cite{ganguli2022predictability}. 
 \add{Unlike conventional AI, which involves inputting data and receiving a decision or analysis, GenAI allows users to iteratively provide initial inputs (e.g., prompts) and receive AI-generated content. This process requires active user involvement in guiding and refining, resulting in varied user experiences based on their prompting strategies, domain knowledge, and GenAI proficiency. 
 Our study corroborates previous findings on usability issues in GenAI, such as challenges in controllability and effective prompting for non-expert users~\cite{zamfirescu2023johnny,wang2023reprompt,wu2022aichains}.} 

\add{More interestingly, we found that} through extended experimentation with GenAI tools, users tend to develop their own prompting strategies after forming their own `mental models'~\cite{carroll1988mental,allen1997mental} of how GenAI interprets prompts and generates outputs. For instance, P12 observed that altering the placement of words in prompts could result in those words receiving more emphasis in content generation, \add{which informed his prompting strategy.} 
Other participants developed more systematic strategies such as `role assignment, task contextualization, task decomposition, and task specification' [P3] and `summarizing complex ideas into one-line prompts' [P9]. \add{This process of hypothesizing and testing reflects an iterative approach to understanding and experimenting with GenAI, based on the user's evolving mental model.} In addition, users also actively seek available learning resources \add{to enhance their GenAI skills}, such as online courses and tutorials shared by influencers on social media, \rev{These inconsistent findings underscore the importance of conducting longitudinal studies to investigate users' strategies for developing prompt design approaches thoroughly.}{highlighting the inherently intertwined nature of the learning and using phases within the human-GenAI co-creation process.}

\section{Practical and Design Implications}

\subsection{Supporting User Agency}

\subsubsection{Design implications for user control}
Our study involved participants with varying GenAI expertise: while some were satisfied with the basic functions of current GenAI tools, others sought affordances that offer granular control to achieve desired outcomes. For example, several participants desired a feature to localize changes to specific parts of an image during text-to-image generation. Thus, in addition to the current efforts in developing new techniques, such as new text-to-image generative models~\cite{zhang2023adding}, it is also important to design affordances that support more fine-grained control. Features that allow users to precisely specify which factors (e.g., color, atmosphere, and texture) to keep while experimenting with others will be highly appreciated. Additionally, participants with non-technical backgrounds expressed a need for more user-friendly interfaces. For instance, many current text-to-video generation tools demand proficiency in mathematics and programming for effective camera movement control. A design that is more intuitive and user-centric is highly sought after for future GenAI tools~\cite{hurtienne2007intuitive}.

\subsubsection{Design implications for feedback mechanisms}

While prior studies~\cite{sundar2023calling,limerick2014agency} pointed out the importance of soliciting user feedback to enhance user agency, our study found that the feedback mechanisms in existing GenAI tools (e.g., the up-/down-vote option in ChatGPT) are perceived as non-intuitive and ineffective. Thus, the design of future GenAI feedback mechanisms could be enhanced in multiple ways: i) improved visibility; ii) providing channels for users to specify and clarify their inputs~\cite{sun2022exploring}, and iii) offering incentives for users to provide feedback~\cite{ouyang2022feedback}.

For example, when users first interact with GenAI tools, it is critical to provide brief, user-friendly guidance that explains the value of feedback and demonstrates how their feedback can lead to improved outcomes~\cite{stojmenovic2014visual}.

\subsection{Preventing Digital Divide}
The digital divide refers to the disparity between individuals who have access to digital technologies and those who do not~\cite{van2006digital}. GenAI, being an emerging technology, has the potential to enlarge this existing disparity. Our study reflected the potential of both the `first-level divide'~\cite{bozkurt2023challenging}, which concerns individuals without access to GenAI and the `second-level divide'~\cite{dwivedi2023so}, which concerns the varying acceptance of GenAI across individuals and cultures. For example, a major challenge facing China-based participants is the limited access to U.S.-based GenAI tools and learning resources. 
Our research advocates for enhancing the accessibility of GenAI and offering training related to GenAI literacy, aiming to ensure equitable access and knowledge distribution to help prevent the digital divide~\cite{fui2023generative}. 

Further, we noted that participants with higher levels of knowledge and skill tend to be more adept at circumventing the limitations of current GenAI tools to maximize their effectiveness. These `power users' often utilize GenAI in more sophisticated or advanced manners~\cite{bughin2018poweruser, sundar2020rise}. Thus, gathering the experiences of power users with GenAI and creating an extensive knowledge base would be highly beneficial. This resource would function as a central reference for addressing frequent questions and resolving issues (e.g.,providing comprehensive tutorials for new users).

\subsection{Implementing Responsible GenAI}
Multiple participants voiced concerns about inputting sensitive information into GenAI tools, due to the unclear usage of data by such tools. \add{Although the U.S. government has recently issued an executive order establishing new standards for AI safety and security~\cite{WhiteHouse2023}, particularly concerning AI-generated content, it notably lacks any reference to multilingual content or translation. Further, GenAI regulations and standards vary greatly across countries, presenting additional challenges for global creative industries.} 
Despite the caution shown by companies in navigating this diverse, evolving regulatory landscape, it is crucial to recognize that employees are 
increasingly depending on GenAI for various work-related activities. Consequently, guiding employees on how to effectively and ethically utilize GenAI is becoming increasingly vital. Companies should establish clear guidelines for GenAI usage and promote GenAI-related professional development, ensuring that employees remain abreast of the latest advancements in the field. \add{Also, it is prudent to uphold proactive transparency regarding GenAI-generated content, especially in multilingual contexts.}

\section{Limitations and Future Work}
Firstly, the majority of our participants were from creative industries. While our study provides valuable insights into perceptions and uses of GenAI in real-world contexts, it is critical to acknowledge its potential lack of representativeness among other populations. Future research should explore more diverse professional domains \add{beyond creative industries} (e.g., academic research~\cite{liebrenz2023scholarly}). Future studies should also employ quantitative methods to improve the generalizability of our findings. Secondly, \add{the recruitment process was influenced by our personal networks and social media platforms, encompassing diverse backgrounds within creative industries.} The voluntary participation in our study may introduce self-selection bias~\cite{robinson2014sampling}, as users who consented might be more active and advanced GenAI users than those who did not participate. This raises questions for future research concerning non-use or reluctance to adopt GenAI~\cite{satchell2009nonuse}. Thirdly, our study focuses on the present state of GenAI. Considering the rapidly evolving nature of GenAI technologies, future research should systematically explore emerging issues and challenges not fully covered in this study, such as GenAI's role in information validation and strategies to overcome hallucination~\cite{zhou2023synthetic}. Finally, non-U.S. participants raised unique issues such as technological and infrastructure barriers in the application of GenAI. \add{Future studies should conduct comparative analyses of GenAI users across various countries to evaluate how regional and cultural factors impact the adoption and utilization of GenAI.}

\section{Conclusion}

\add{In this study, through the lens of a holistic ``learning, using, assessing'' framework, we explored how users perceive and utilize GenAI technologies `in the wild,' leading to a number of intriguing findings: users hold varied and nuanced perspectives on GenAI's unique characteristics, particularly its creativity; they interact with GenAI through a dynamic, evolving human-GenAI co-creation process; they encounter various challenges beyond GenAI's functional limitations; and they actively exert their agency to overcome many of these challenges. Our study highlights the intricate synergy between humans and AI in this GenAI era, opening up several promising avenues for future studies.}

\begin{acks}
We thank our participants for sharing their thoughts and experiences. We also thank the anonymous reviewers for their valuable feedback. This work is supported by the National Science Foundation under Grant No. 2212323, 1951729, and 1953893.
\end{acks}



\bibliographystyle{ACM-Reference-Format}
\bibliography{bibs/Agency,bibs/GenAI,bibs/HAICollaboration,bibs/Creativity,bibs/InfoValidation,bibs/Limitation,bibs/Revisions}

\end{document}